\documentclass[12pt]{article}
\usepackage{amsfonts,amsmath,amssymb,mathrsfs}

\title{A Relativistic Version of the\\ Ghirardi--Rimini--Weber
       Model}
\author{ 
Roderich Tumulka\footnote{Dipartimento di Fisica dell'Universit\`a
    di Genova and INFN sezione di Genova,
    Via Dodecaneso 33, 16146 Genova, 
    Italy. E-mail: tumulka@mathematik.uni-muenchen.de}
}
\date{August 16, 2006}

\newcommand{\RRR}{\mathbb{R}}
\newcommand{\CCC}{\mathbb{C}}
\newcommand{\vx}{\boldsymbol{x}}
\newcommand{\vy}{\boldsymbol{y}}
\newcommand{\vr}{\boldsymbol{r}}
\newcommand{\vz}{{\boldsymbol{z}}}
\newcommand{\vX}{{\boldsymbol{X}}}
\newcommand{\vY}{{\boldsymbol{Y}}}
\newcommand{\I}{i}
\newcommand{\1}{1}

\newcommand{\prob}{\mathrm{Prob}}
\newcommand{\fut}{\mathscr{F}} 
\newcommand{\past}{\mathscr{P}} 
\newcommand{\sph}{\mathscr{H}} 
\newcommand{\st}{\mathscr{M}} 
\newcommand{\D}{\mathscr{D}} 
\newcommand{\tdist}{\text{t-dist}} 
\newcommand{\sdist}{\text{s-dist}} 

\begin{document}\maketitle\sloppy
\begin{abstract}
  Carrying out a research program outlined by John~S.~Bell in 1987, we
  arrive at a relativistic version of the Ghirardi--Rimini--Weber
  (GRW) model of spontaneous wavefunction collapse. The GRW model
  was proposed as a solution of the measurement problem of 
  quantum mechanics and involves
  a stochastic and nonlinear modification of the Schr\"odinger equation.
  It deviates very little from the Schr\"odinger equation for microscopic 
  systems but efficiently suppresses, for macroscopic systems, 
  superpositions of macroscopically
  different states. As suggested by
  Bell, we take the primitive ontology, or local beables, of our model
  to be a discrete set of space-time points, at which the collapses
  are centered. This set is random with distribution determined by the
  initial wavefunction.  Our model is nonlocal and violates Bell's
  inequality though it does not make use of a preferred slicing of
  space-time or any other sort of synchronization of spacelike
  separated points.  Like the GRW model, it reproduces the quantum
  probabilities in all cases presently testable, though it entails
  deviations from the quantum formalism that are in principle
  testable.  Our model works in Minkowski space-time as well as in
  (well-behaved) curved background space-times.

\medskip

\end{abstract}

\section{Introduction}

\begin{quotation}
  \textit{\ldots I am particularly struck by the fact that the [GRW]
  model is as Lorentz invariant as it could be in the nonrelativistic
  version. It takes away the ground of my fear that any exact
  formulation of quantum mechanics must conflict with fundamental
  Lorentz invariance.} \hfill J.~S.~Bell \cite{Belljumps}
\end{quotation}

\noindent In 1986, Ghirardi, Rimini, and Weber (GRW) proposed a model of spontaneous wavefunction collapse \cite{grw} based on a stochastic and nonlinear modification of the Schr\"odinger equation. When combined with a clear ontology, the GRW model turns quantum mechanics into a completely coherent theory. It resolves all paradoxes of quantum mechanics, in particular the measurement problem, and accounts for all phenomena of quantum mechanics in terms of an objective reality governed by mathematical laws. However, the GRW model is nonrelativistic.

In \cite{Belljumps} and again in \cite{Bellexact},
John~S.~Bell emphasized that the GRW model has a
property---multi-time translation invariance---that can be regarded as
a nonrelativistic surrogate of Lorentz invariance.  This fact suggests
that the biggest difficulty one would expect with turning a
nonrelativistic theory into a relativistic one---the difficulty caused
by the lack of a temporal ordering of spacelike separated events---is
absent in the GRW model right from the start.  We find this suggestion
to be correct, and can indeed specify a relativistic version of the
GRW model.  We proceed along the lines of Bell's suggestions; in
particular, we do not use a \emph{continuous} spontaneous collapse
model (corresponding to a diffusion process in Hilbert space and known
as continuous spontaneous localization, or CSL), but rather a
\emph{discrete} one corresponding to a jump process in Hilbert space.
Furthermore, we follow Bell in taking as the primitive ontology, or
local beables, of the model the space-time points where the collapses
are centered \cite{Belljumps,Bellexact,Kent89}.  ``A piece of matter then is a galaxy of such events''
\cite{Belljumps}.  We will call these points ``flashes.''

For a recent overview of spontaneous collapse models, see \cite{overview}.
We regard our model as a step towards one possible explanation of the
probability rules of quantum theory in the relativistic realm.  This
realm differs in two ways from that of nonrelativistic quantum
mechanics, to which the GRW model applies: in the requirement of
Lorentz symmetry, and in the phenomenon of particle creation and
annihilation typical of quantum field theory. Here we will not be
concerned with the latter, but focus on covariance, and thus on
relativistic quantum mechanics; correspondingly, what we shall mean by
``relativistic'' is ``Lorentz invariant,'' or its analogue in curved
space-time.

Our relativistic model is surprisingly similar to the original GRW
model, which it approaches in the nonrelativistic limit.  Its
structure is in no way more complicated than that of the GRW model in
Bell's flash-based version.  The two models have the following
features in common: (i)~the only objects in the universe (beyond the
given space-time geometry) are the wavefunction and the flashes;
(ii)~two new constants of nature are needed, the collapse rate
$1/\tau$ per particle and the width $a$ of the localization;
(iii)~time reversal invariance is broken, while (in flat space-time)
rotation, space translation, time translation, parity, and gauge
invariance are obeyed; (iv)~the dynamics is intrinsically
stochastic.

Our model is based on relativistic quantum mechanics of $N$ particles.
The question of identical particles we plan to address in a separate
work; here we shall avoid this question and base our considerations on
the quantum mechanics of distinguishable particles.  We shall use the
letter $i$ to denote (apart from $\sqrt{-1}$) the particle types, $i
\in \{1,\ldots,N\}$, and $Q_i$ for the set of flashes belonging to
$i$; the elements of this set are timelike separated from each other.

The wavefunction is a multi-time wavefunction, i.e., it is defined on
the Cartesian product of $N$ copies of space-time.  We use the Dirac
equation as the relativistic version of the Schr\"odinger equation
determining the evolution of the wavefunction apart from the collapses
(but we will mostly not worry whether the wavefunction lies in the positive
energy subspace, except in 
Section~\ref{sec:positron}).  More precisely, we use the multi-time formalism
with $N$ Dirac equations.  For the consistency of this set of
equations, we cannot have interaction potentials.  To avoid discussing
the question of interaction in relativistic quantum mechanics, we will
assume non-interacting particles.  Interaction can presumably be
included by allowing for particle creation and annihilation, which
however is beyond the scope of this paper.  In any case, the
difficulty of including interaction that we encounter here does not
stem from the spontaneous collapses but rather from the mathematics of
multi-time equations, and is thus encountered by every kind of
relativistic quantum mechanics.

We will give three equivalent descriptions of our model. In one of
them, we refer to an arbitrary slicing (foliation) of space-time into
spacelike surfaces and obtain a Markov process for the temporal
evolution (relative to this slicing) of wavefunction and flashes;
since this picture is not manifestly covariant, we postpone it to the
end.  We will begin instead with an iterative construction of the
flashes, and take this to be the definition of our model.  In another
description, we provide a formula for the joint distribution of the
flashes on space-time in terms of the initial wavefunction.

This paper is organized as follows. In Section~\ref{sec:grw} we recall
the definition of the GRW model. In Section~\ref{sec:model} we define
our relativistic variant. In Section~\ref{sec:formulas} we compute
joint distributions of the flashes and flash rates.  In
Section~\ref{sec:temporal} we reformulate the model in terms of a
temporal evolution relative to an arbitrary spacelike slicing of
space-time.  In Section~\ref{sec:low} we show that the low velocity
limit of our model is the GRW model.  In Section~\ref{sec:predictions}
we discuss some predictions of our model.  Finally, in
Section~\ref{sec:conclusions} we conclude by comparing it to other
models in the literature.

\section{The GRW Model}\label{sec:grw}

\subsection{Definition}

We briefly recall the GRW model, following Bell's description
\cite{Belljumps}.  The wavefunction $\Psi = \Psi(\vr_1, \ldots,
\vr_N,t)$ evolves unitarily between the collapses.  At the time $T$
when a flash of type $I\in \{1, \ldots, N\}$ occurs, at location $\vX
\in \RRR^3$, the wavefunction collapses according to
\begin{equation}\label{grwcollapse}
  \Psi(\vr_1, \ldots, \vr_N, T+) = \frac{j(\vr_I-\vX) \, \Psi(\vr_1,
  \ldots, \vr_N, T-)}{\rho^{1/2}_I (\vX, T-)}
\end{equation}
where the jump factor is a Gaussian
\begin{equation}\label{grwjdef}
  j(\vr) = K \, \exp \Bigl(-\frac{\vr^2}{2a^2}  \Bigr)
\end{equation}
whose width $a$ is a new constant of nature, of order of magnitude
$10^{-7} \, \text{m}$, and the normalization constant $K$ is chosen so
that
\begin{equation}\label{grwKdef}
  \int_{\RRR^3} d^3 \vr \; |j(\vr)|^2 = 1.
\end{equation}
Furthermore,
\begin{equation}\label{grwrhodef}
  \rho_i(\vx,t) = \int_{\RRR^{3N}} d^3\vr_1 \cdots d^3\vr_N \: \bigl|
  j(\vr_i-\vx) \, \Psi(\vr_1, \ldots, \vr_N,t) \bigr|^2,
\end{equation}
so that the collapsed wavefunction in \eqref{grwcollapse} is
normalized.  The rate for a collapse of type $i$ to occur in the
volume element $d^3\vy$ is
\begin{equation}\label{grwcollapserate}
  \frac{1}{\tau} \, \rho_i(\vy) \, d^3\vy,
\end{equation}
where $\tau$ is another new constant of nature, of order of magnitude
$10^{15} \, \text{sec}$.  To put this differently,
\begin{equation}\label{grwcollapserate2}
  \prob \Bigl( Q_i\cap [t,t+dt]\times \RRR^3 = \{Y\},Y \in [t,t+dt]
  \times d^3\vy \Big| \Psi_t \Bigr) = \frac{dt\, d^3\vy}{\tau}
  \bigl\langle \Psi_t \big| \hat\jmath_i(\vy)^2 \big| \Psi_t
  \bigr\rangle,
\end{equation}
where $\hat\jmath_i(\vy)$ is the self-adjoint ``collapse'' operator
that multiplies by the function $j(\vr_i-\vy)$, and $Q_i$ is the set of all
flashes with label $i$ (a subset of space-time).

\subsection{Another Definition}\label{sec:another}

We now give a second, equivalent, formulation of the GRW model that
is closer to the way we will formulate the definition of our
relativistic model.  A first change consists in that, rather than
having the wavefunction change discontinuously at some time $T$, we
will speak of two different wavefunctions, one representing the
situation before collapse and the other the situation after (or better,
one uncollapsed and one collapsed), and
extend both wavefunctions to all times, future and past, using the
unitary, collapse-free evolution.  Hence, in this terminology and
notation, wavefunctions never collapse.  If that seems paradoxical, we
emphasize that the goal is to define the distribution of the random
sets $Q_i$, as it is only the flashes that chairs, tables, and
observers are made of, and whatever mathematical formulation leads to
the right distribution is allowable.

Assume, as we will do in the relativistic model, that the $N$
particles do not interact, that is, that the Hamiltonian is of the
form $H = H_1 + \ldots + H_N$ where $H_i$ acts only on the $i$-th
coordinate of the wavefunction---so that $H_i$ commutes with $H_{j}$,
$i\neq j$.  This allows us to define the Schr\"odinger evolution also
for a multi-time wavefunction
\begin{equation}\label{multitime}
  \Psi(\vr_1,t_1, \ldots, \vr_N,t_N) = e^{-\I H_1t_1/\hbar} \cdots
  e^{-\I H_Nt_N/\hbar} \Psi(\vr_1,0,\ldots,\vr_N,0).
\end{equation}

The building block of the reformulation of the GRW model is the
following procedure for obtaining, from given time values $T_1,
\ldots, T_N$ (later taken to be times of flashes) and a given
wavefunction $\Psi$ on $\RRR^{4N}$, new random time values $T_1',
\ldots, T_N'$ (later taken to be times of subsequent flashes),
associated random locations $\vY_1, \ldots, \vY_N \in \RRR^3$, and a
new (collapsed) wavefunction $\Phi$ on $\RRR^{4N}$: Let $\Delta T_1,
\ldots, \Delta T_N$ be independent, exponentially distributed random
variables with expectation $\tau$, and set $T_i' = T_i + \Delta T_i$.
The joint distribution of the $\vY_i$,
\begin{equation}\label{grwYdistr}
  \prob \bigl(\vY_1 \in d^3\vy_1, \ldots, \vY_N\in d^3 \vy_N \bigr) =
  \rho(\vy_1, \ldots,\vy_N) \, d^3 \vy_1 \cdots d^3 \vy_N,
\end{equation}
has density $\rho: \RRR^{3N} \to \RRR$ defined by
\begin{equation}\label{grwmultitimerhodef}
  \rho(\vy_1, \ldots, \vy_N) = \int_{\RRR^{3N}} d^3\vz_1 \cdots d^3
  \vz_N \: \bigl|j(\vz_1- \vy_1) \cdots j(\vz_N- \vy_N) \, \Psi(\vz_1,
  T_1', \ldots, \vz_N, T_N') \bigr|^2.
\end{equation}
Now define, for all $(\vz_1, \ldots, \vz_N) \in \RRR^{3N}$,
\begin{equation}
  \Phi(\vz_1, T_1', \ldots, \vz_N, T_N') = \frac{j(\vz_1-\vY_1) \cdots
  j(\vz_N-\vY_N) \, \Psi(\vz_1,T_1', \ldots,
  \vz_N,T_N')}{\rho^{1/2}(\vY_1, \ldots, \vY_N)}
\end{equation}
and extend $\Phi$ to all other times values by unitary multi-time
evolution such as \eqref{multitime}.

The random sets $Q_i$ of flashes are obtained by iterating this
procedure. We start with $T_1 =0 , \ldots, T_N=0$ and the initial
wavefunction, obtain the first flash $(T_i',\vY_i)$ in every $Q_i$,
and then take the times of these flashes and the new wavefunction as
the input of the next round of the procedure.  It can be shown that
this definition is equivalent to the definition given in the previous
subsection as it leads to the same random sets $Q_i$.
We also obtain in this way a sequence of wave functions,
each on $\RRR^{4N}$, each a solution to \eqref{multitime}, and
each associated with a set of $N$ flashes.

\section{The Relativistic Model}\label{sec:model}

In this section we define the relativistic model.  We will use a
rather abstract formulation that has the advantages of being concise,
simple, manifestly consistent, and manifestly covariant.

\subsection{The Dirac Equation}

We begin with recalling some relevant aspects of the Dirac equation
and introducing some notation along the way.  We generally use the
letters $x,y,\ldots,X,Y,\ldots$ to denote space-time points, where
capital letters usually stand for random space-time points.

As the particles are not interacting, our model does not require that
they live in the same space-time; instead, we may have $N$ space-time
manifolds $\st_1, \ldots,\st_N$ (which may be flat or curved), such
that the wavefunction is a function on $\st_1 \times \cdots \times
\st_N = \prod_i \st_i$, and the set $Q_i$ of flashes of type $i$ is a
discrete subset of $\st_i$. We find that this greater mathematical
generality, though physically unnecessary, facilitates the
mathematical treatment of the model.

The evolution of the wavefunction $\Psi = \Psi(x_1, \ldots, x_N)$ is
determined, apart from the collapses, by the Dirac equation
\begin{equation}\label{Dirac}
  \I\hbar \gamma_i^\mu \bigl( \nabla_{i,\mu} - \tfrac{\I e_i}{\hbar}
  A_{i,\mu}(x_i) \bigr) \Psi = m_i \Psi.
\end{equation}
Here, $m_i$ and $e_i$ are mass and charge of particle $i$, $\nabla_i$
is the (covariant) derivative on $\st_i$, and $A_i$ the
electromagnetic vector potential on $\st_i$. $\Psi$ takes values in
$(\CCC^4)^{\otimes N}$ or, in the case of curved space-times, is a
cross-section of the vector bundle
\begin{equation}
  \D = \bigcup_{x_1 \in \st_1, \ldots, x_N \in \st_N}
  \D_{(x_1,\ldots,x_N)} = \bigcup_{x_1 \in \st_1, \ldots, x_N \in
  \st_N} \D_{1,x_1} \otimes \cdots \otimes \D_{N,x_N},
\end{equation}
where $\D_i$ is the bundle of Dirac spin-spaces associated with
$\st_i$.

As in Section~\ref{sec:another}, we will, throughout
Section~\ref{sec:model}, take the wavefunctions to always obey the
collapse-free evolution, this time given by the multi-time Dirac
equation \eqref{Dirac}, and to be defined on all of $\prod_i \st_i$.
In particular, at any collapse we will consider two wavefunctions
rather than one that changes discontinuously.

A remark is necessary on consistency of the multi-time formalism.
Multi-time equations are not always consistent; what can go wrong is
that the propagator for the $i$-th time coordinate (relative to some
coordinate system) fails to commute with the $j$-th propagator, $j
\neq i$, in which case it is impossible to find, for every initial
wavefunction, a wavefunction on $\prod_i \st_i$ solving all $N$
evolution equations simultaneously.  The set of equations
\eqref{Dirac}, however, is consistent because the metric and vector
potential acting on the coordinate $x_i$ do not depend on the other
coordinates $x_j$, $j\neq i$.

We now briefly explain how the Dirac equation \eqref{Dirac} defines
unitary propagators on suitable Hilbert spaces. Consider first the
one-particle Dirac equation
\begin{equation}\label{1Dirac}
  \I\hbar \gamma^\mu \bigl( \nabla_{\mu} - \tfrac{\I e}{\hbar} A_{\mu}
  \bigr) \Psi = m \Psi.
\end{equation}
We need and assume from now on that there are no closed timelike
curves.  With every spacelike surface $\Sigma$ is associated a Hilbert
space $L^2(\Sigma)$, and with any two spacelike Cauchy\footnote{A
Cauchy surface is a surface intersected by every complete timelike
curve.}  surfaces $\Sigma$ and $\Sigma'$ is associated a unitary
propagator $\hat U_\Sigma^{\Sigma'} : L^2( \Sigma) \to L^2(\Sigma')$
as follows.  $L^2(\Sigma)$ contains cross-sections of the bundle $\D$
of Dirac spin spaces (which on Minkowski space-time are just $\CCC^4$)
restricted to $\Sigma$ and is endowed with the scalar product
\begin{equation}
  \langle{\Psi}|{\Phi}\rangle = \int_\Sigma d^3x \: \overline\Psi(x)
  \, n_\mu(x) \, \gamma^\mu \, \Phi(x)
\end{equation}
where $n_\mu(x)$ is the (future-directed) unit normal vector on $\Sigma$ at $x$, and the
volume measure $d^3x$ is the one arising from the Riemann metric on
$\Sigma$. For simplicity, we will often write
\begin{equation}
  |\Psi(x)|^2 = \overline\Psi(x) \, n_\mu(x) \, \gamma^\mu \, \Psi(x)
\end{equation}
when it is clear which surface $\Sigma$ we are considering.  Unitarity
of the propagator $\hat U_\Sigma^{\Sigma'}$ follows from the
continuity equation
\begin{equation}
  \nabla_\mu \Bigl( \overline\Psi \gamma^\mu \Psi \Bigr) = 0,
\end{equation}
which is a consequence of \eqref{1Dirac}.  Of course, $\hat
U_{\Sigma'}^{\Sigma''} \, \hat U_{\Sigma}^{\Sigma'} = \hat
U_\Sigma^{\Sigma''}$, and $\hat U_\Sigma^\Sigma$ is the identity
operator on $L^2(\Sigma)$.  For multi-time wavefunctions, we will only
consider spacelike surfaces of product form, $\Sigma_1 \times \cdots
\times \Sigma_N$. Then, the associated Hilbert space coincides with
the tensor product of the Hilbert spaces associated with the
$\Sigma_i$, and the unitary propagator between two such spaces
coincides with the tensor product of the unitary propagators for each
$i$.

\subsection{Notation}

We need some more notation.  Let $\fut(x)$ denote the future of $x$,
i.e., the future light cone and its interior including $x$ itself, and
$\past(x)$ the past of $x$.  When $S$ is a subset of space-time we
write $\fut(S)$ (``the future of $S$'') for $\bigcup_{x\in S} \fut(x)$
and $\past(S)$ (``the past of $S$'') for $\bigcup_{x \in S} \past(x)$.
For $y \in \fut(x)$, let $\tdist(y,x)$ be the timelike distance of $y$
from $x$, i.e., the supremum of the lengths of all timelike curves
connecting $x$ to $y$; for Minkowski space-time,
\begin{equation}
  \tdist(y,x) = \bigl((y^\mu-x^\mu)(y_\mu -x_\mu) \bigr)^{1/2}.
\end{equation}
For $r\geq 0$, let
\begin{equation}
  \sph_r(x) = \{y \in \fut(x): \tdist(y,x) =r\}
\end{equation}
be the future surface of timelike distance $r$ from $x$, or the
$r$-\emph{hyperboloid}; for $r=0$ this is the future light cone. For
$y \in \fut(x)$, we also write $\sph(y,x) = \sph_{\tdist(y,x)}(x)$ for
the hyperboloid centered at $x$ containing $y$.  If $\Sigma$ is a
spacelike surface and $x,y \in \Sigma$, we write $\sdist_\Sigma(x,y)$
for the spacelike distance from $x$ to $y$ along $\Sigma$, i.e., the
infimum of the Riemannian lengths of all curves in $\Sigma$ connecting
$x$ to $y$.  The speed of light is denoted by $c$.

\subsection{Law of the Flashes}

We are now prepared to write down the stochastic law of the flashes.
The building block is the following procedure for obtaining, from given
flashes $X_1, \ldots, X_N$ and a (normalized) wavefunction $\Psi$
obeying \eqref{Dirac}, new flashes $Y_1, \ldots, Y_N$ and a new
wavefunction $\Phi$ obeying \eqref{Dirac}.

Let $\Delta T_1, \ldots, \Delta T_N$ be independent, exponentially
distributed random variables with expectation $\tau$. Choose $(Y_1,
\ldots, Y_N)$ at random from $\Sigma_1 \times \cdots \times \Sigma_N =
\prod_i \Sigma_i$, where $\Sigma_i = \sph_{c\Delta T_i}(X_i)$, with
distribution
\begin{equation}\label{Ydistr}
  \prob \bigl(Y_1 \in d^3y_1, \ldots, Y_N\in d^3 y_N \bigr) =
  \rho(y_1, \ldots,y_N) \, d^3 y_1 \cdots d^3 y_N
\end{equation}
as follows. The volume of $d^3y_i$ is computed using the Riemann
metric on $\Sigma_i$; the distribution density $\rho: \prod_i \Sigma_i
\to \RRR$ is defined by
\begin{equation}\label{rhodef}
  \rho(y_1, \ldots, y_N) = \int_{\prod_i \Sigma_i} d^3z_1 \cdots d^3 z_N
  \: \bigl|j_{\Sigma_1}(y_1, z_1) \cdots j_{\Sigma_N}(y_N, z_N) \,
  \Psi(z_1, \ldots, z_N) \bigr|^2;
\end{equation}
and, for any spacelike surface $\Sigma$, the jump factor
$j_\Sigma: \Sigma \times \Sigma \to \RRR$ is defined by
\begin{equation}\label{jdef}
  j_\Sigma(y, z) = K_\Sigma(z) \, \exp\biggl( - \frac{ \sdist^2_{\Sigma}
  (y,z)}{2a^2} \biggr)
\end{equation}
with $K_\Sigma(z)$ chosen so that 
\begin{equation}\label{jnormalized}
  \int_\Sigma d^3 y \; |j_\Sigma(y, z)|^2 = 1.
\end{equation}
Now define $\Phi$ on $\prod_i \Sigma_i$ by
\begin{equation}\label{Phidef}
  \Phi(z_1, \ldots, z_N) = \frac{j_{\Sigma_1}(Y_1, z_1) \cdots
  j_{\Sigma_N}(Y_N, z_N) \, \Psi(z_1, \ldots,
  z_N)}{\rho^{1/2}(Y_1,\ldots,Y_N)},
\end{equation}
and define it on the remainder of $\prod_i \st_i$ by extending via
\eqref{Dirac}.

To the extent that the hyperboloids $\Sigma_i = \sph_{c\Delta
T_i}(X_i)$ are Cauchy surfaces (see remarks in the subsequent
subsection), $\Phi$ is uniquely determined on $\prod_i \st_i$ from
initial data on $\prod_i \Sigma_i$ (and normalized due to
\eqref{rhodef}), $\Psi$ is normalized on $\prod_i \Sigma_i$, and, by
\eqref{jnormalized}, $\rho$ is normalized in the sense
\begin{equation}
  \int_{\prod_i \Sigma_i} d^3y_1 \cdots d^3y_N \: \rho(y_1,\ldots,
  y_N) =1.
\end{equation}
Note also that $Y_i \in \fut(X_i)$.

Our relativistic model is defined by iterating this procedure. As
initial data, specify a wavefunction $\Psi=\Psi^{0}$ and one flash
$X_i^{0}$ of every type $i$. Apply the above procedure to obtain
$\Psi^{1} = \Phi$ and a new flash $X_i^{1}=Y_i$ of every type.  Repeat
the procedure with $\Psi = \Psi^{1}$ and $X_i = X_i^{1}$, and so
on. In this way, obtain for every type $i$ a random sequence of
flashes, $Q_i = \{X_i^{0}, X_i^{1}, X_i^{2}, \ldots\}$.

Concerning the initial flashes of the universe, it seems an idea worth considering 
that the Big Bang, i.e., the initial singularity of the space-time geometry, 
is the space-time location of the initial flash for each $i\in \{1,\ldots, N\}$.
This could have the status of a law, and would remove the arbitrariness of the initial flashes. Of course, the mathematics of the model works with any choice
of the initial flashes.

\subsection{Hyperboloids and Cauchy Surfaces}\label{sec:cauchy}

A hyperboloid need not be a Cauchy surface in $\st_i$; in fact, it
never is in Minkowski space-time,\footnote{To see this, consider the
following example of a complete timelike curve $x(t)$ that does not
intersect the unit hyperboloid $\sph_1(0,0,0,0)$: the uniformly
accelerated curve $x(t) = (ct,0,0,\sqrt{1+c^2 t^2})$.}  though it is
``almost Cauchy'' in the sense that most complete timelike
curves do intersect a given hyperboloid: the ones avoiding the
hyperboloid accelerate to the speed of light.  However, it is not
necessary for our purposes that the hyperboloids be Cauchy surfaces.
It is sufficient that, for a single particle,
\begin{equation}\label{evolution}
\begin{array}{l}
  \text{the Dirac equation defines a unitary evolution operator}\\
  U_{\Sigma}^{\sph}: L^2(\Sigma) \to L^2(\sph)  \text{ for every
  Cauchy surface }\Sigma \\ \text{and every 
  hyperboloid }\sph.
\end{array}
\end{equation}
We conjecture, but do not have a proof, that this is the case
in Minkowski space-time for a large class of vector potentials $A_{\mu}$.
We hope to be able to provide a proof in a future work.
Here is a partial argument: To show that the evolution is unitary requires
to show that it is onto and norm-preserving. To show the latter it should suffice to 
show that, for all wavefunctions $\psi$ from a dense subspace of 
$L^2(\Sigma)$, no more than a set of
measure zero of flow lines of the $|\psi|^2$ distribution (Bohmian
trajectories) avoid the hyperboloid.
This presumably amounts to the vanishing of the probability flux into
the future null infinity.  This could be expected to be the case
for $\psi$ from the form domain of the Hamiltonian,
because to accelerate a positive amount of Bohmian trajectories to the
speed of light should require, in some sense, an infinite amount of
energy, while such $\psi$ has finite energy expectation.  Indeed this
conclusion apparently follows, for static
electromagnetic fields tending to zero fast enough at infinity and
for a suitable class of $\psi$'s, from the flux-across-surfaces
theorem \cite{peter} using the global existence of Bohmian
trajectories \cite{bdex}.

How about other space-times than Minkowski?
In some space-times, hyperboloids actually are Cauchy surfaces and so
the problem is absent; for example, think of Minkowski space-time modulo
a spacelike 3-lattice.  Relevant conditions for this case may be emptiness
of the future null infinity and absence of future singularities.  The
future null infinity could be expected to be empty when ``space has
finite volume growing not too quickly,'' such as in
Minkowski space modulo a spacelike 3-lattice. 
In space-times in which, as in Minkowski space-time, 
hyperboloids are not Cauchy surfaces, a relevant condition for
\eqref{evolution} seems to be that, for all $\psi$ from a dense subspace,
the probability flux into the future null infinity vanishes.

For particles with zero rest mass, such as photons and
gravitons, one would expect that there typically is a positive
probability flux into the future null infinity.\footnote{This remark
has been kindly pointed out to me by Fay Dowker.}  One way of avoiding
this problem would be to postulate that flashes are associated only
with massive particles, and to have the arguments of the wavefunction
corresponding to massless particles get integrated out in
\eqref{rhodef} along an arbitrary Cauchy surface.

\subsection{Nonlocality}

The model is nonlocal. This is manifest in \eqref{Ydistr} in that the
joint distribution of the flashes $Y_i$ and $Y_j$, $j \neq i$, does
not factorize. One can easily find situations in which the events
$Y_i$ and $Y_j$ (now taken to lie in the same space-time manifold) are
spacelike separated, and still the distribution of $Y_i$ depends on
the realization of $Y_j$; or, of course, we may view it the other way
round: that the distribution of $Y_j$ depends on the realization of
$Y_i$. 

\subsection{Generations}

Since the flashes were constructed here together in \emph{generations}, 
i.e., groups $Y_1, \ldots, Y_N$ of $N$ flashes, one may wonder whether
the theory presupposes, or provides, some structure beyond the
mere flashes and their labels, a structure defining which flashes
belong to the same generation. Such a structure would seem against
the spirit of relativity, as much perhaps as a preferred slicing of
space-time into hypersurfaces, since it would define, for two 
spacelike separated flashes, which is earlier and which is later,
if they belong to different generations.

In fact, however, the theory neither presupposes
nor provides a grouping of the flashes into generations. Rather, one could
take as the initial flashes a combination of some of the first-generation
flashes $X_i$ and some of the second-generation flashes $Y_i$,
and obtain the same (conditional) distribution of the future flashes.
This will become manifest in the reformulation of the law of the flashes of 
Section~\ref{sec:firstn}.

\subsection{Negative Energy Contributions}\label{sec:positron}

In the context of the Dirac equation, one often considers 
wavefunctions with negative energy as unphysical, more precisely 
those wavefunctions containing, for at least one particle, at least 
some contribution from the negative-energy subspace of the 1-particle
Dirac Hamiltonian. Such wavefunctions we will call ``non-positive'' for short, 
and the other wavefunctions, containing exclusively positive-energy contributions, 
``positive.'' \emph{The collapse accompanying a flash will generically map 
a positive wavefunction to a non-positive one.} This fact poses a
problem if we want to take the model seriously, and in particular if we 
want to extend it to quantum field theories incorporating anti-particles.

One might conclude that the law for collapsing the wave function
should be so modified that the collapsed wavefunction remains positive,
for example by simply projecting the wavefunction, on top of the usual 
multiplication by a Gaussian as in \eqref{Phidef}, to the space
of positive wavefunctions. In formulas, if $P_+$ denotes this projection,
and if we abbreviate the collapse law \eqref{Phidef} as 
\[
  \Phi \propto j \Psi\,,
\]
the modified collapse law corresponds to
\begin{equation}\label{modPhidef}
  \Phi \propto P_+ (j\Psi)\,.
\end{equation}
Since $P_+$ is the same operator 
as convolution with a suitable function whose width is small 
(of the order of magnitude of the Compton wavelength), and thus 
does not change the $|\Phi|^2$ distribution a lot, replacement of \eqref{Phidef} with
\eqref{modPhidef} would perhaps entail
merely a small change in the distribution of the flashes.

Alternatively, one might conclude that the spontaneous collapses can
lead to spontaneous pair creation, since non-positive wavefunctions
would appear to have something to do with anti-particle states. This effect
might entail observable deviations from quantum mechanics. For
calculating concrete predictions, it would seem necessary to first precisely formulate
a version of the model that incorporates particle creation and uses wavefunctions
from Fock space, but one might guess already from the present model 
that the probability of pair production at a collapse could be of the order
of magnitude of $\|(1-P_+)(j\Psi)\|^2/\|j\Psi\|^2$.

\section{Distribution Formulas}\label{sec:formulas}

In the previous section we have defined, in a rather abstract way, the
distribution of the flashes in space-time. We will find it helpful to
have further expressions for this distribution; such expressions we
provide in this section.

\subsection{The First $n$ Flashes}\label{sec:firstn}

It is now straightforward, though somewhat tedious, to derive an
explicit formula for the joint distribution of the first $n_i$ flashes
of type $i$, given the ``initial'' flashes $X_1^0, \ldots, X_N^0$. 
For any choice of spacelike (Cauchy) surfaces
$\Sigma_i^0$, we can express the distribution in terms of the initial
wavefunction $\Psi^0$, restricted to the Cartesian product of the
$\Sigma_i^0$, in the form
\begin{equation}\label{ndistr}
  \prob \Bigl( X_i^k \in d^4 x_i^k; i=1,\ldots,N; k= 1, \ldots, n_i
  \Bigr) = \biggl\langle \Psi^0 \bigg| E^{(\vec{n})} \biggl(\prod_{i=1}^N
  \prod_{k=1}^{n_i} d^4x_i^k \biggr) \bigg| \Psi^0 \biggr\rangle,
\end{equation}
where the scalar product is taken in $\otimes_i L^2(\Sigma_i^0)$.
Here, $\vec{n}= (n_1, \ldots, n_N)$, and 
$E^{(\vec{n})}$ is a positive-operator-valued measure (POVM) on
$\prod_i \st_i^{n_i}$. It is of a product form,
\begin{equation}
  E^{(\vec{n})} \biggl(\prod_{i=1}^N \prod_{k=1}^{n_i} d^4x_i^k \biggr) =
  \bigotimes_{i=1}^N E_{i,X_i^0}^{(n_i)} \biggl( \prod_{k=1}^{n_i} d^4x_i^k
  \biggr),
\end{equation}
where $E_i^{(n)} = E_{i,x}^{(n)}$ is a POVM on $\st_i^n$, defined on
$L^2(\Sigma_i^0)$ by
\begin{equation}\label{Eidef}
\begin{split}
   E_{i,x_i^0}^{(n)} & \biggl( \prod_{k=1}^n d^4x_i^k \biggr) =
  \biggl( \prod_{k=1}^n d^4 x_i^k \; \1_{\fut(x_i^{k-1})} (x_i^k)
  \biggr) \: \times \\ \times & \: \frac{1}{(c\tau)^{n}} \exp\biggl( -
  \frac{1}{c\tau} \sum_{k=1}^n \tdist(x_i^k,x_i^{k-1}) \biggr) \:
  \hat\jmath_i^1 \, \hat\jmath_i^2 \cdots \hat \jmath_i^n \,
  \hat\jmath_i^n \cdots \hat\jmath_i^2 \, \hat\jmath_i^1
\end{split}
\end{equation}
with $1_B$ the indicator function of the set $B$ and $\hat\jmath_i^k$
the self-adjoint ``collapse'' operator defined on $L^2(\Sigma_i^0)$ as follows. Let, for any
spacelike surface $\Sigma$ and $x \in \Sigma$, $\hat \jmath_\Sigma
(x)$ be the multiplication operator on $L^2(\Sigma)$ that multiplies
by the function $j_\Sigma(x, \cdot)$ defined in \eqref{jdef}. Then
\begin{equation}\label{jhatdef}
  \hat\jmath_i^k =  \hat
  U_{\sph(x_i^k, x_i^{k-1})}^{\Sigma_i^0} \: \hat \jmath _{\sph(x_i^k,
  x_i^{k-1})} (x_i^k) \: \hat U^{\sph(x_i^k, x_i^{k-1})}_{\Sigma_i^0} \,.
\end{equation}
In our notation
$E_{i,x}^{(n)}$ we conceal the (uninteresting) dependence on the
choice of $\Sigma_i^0$. Normalization of
$E_i^{(n)}$, i.e., $E_i^{(n)}(\st_i^n)=\hat{1}$, can also directly 
be seen from the fact that
\begin{equation}
  E_{i}^{(n+1)} \biggl( \prod_{k=1}^n d^4x_i^k \times \fut(x_i^n)
  \biggr) = E_{i}^{(n)} \biggl( \prod_{k=1}^n d^4x_i^k \biggr),
\end{equation}
which in turn follows from
\begin{equation}
  \int_{\fut(x)} d^4y \: \frac{1}{c\tau} \exp \Bigl( -\frac{1}{c\tau}
  \tdist(y,x) \Bigr) \: \hat\jmath_{\sph(y,x)}(y)^2 = \hat{1}.
\end{equation}

Eq.~\eqref{ndistr} is, in effect, a Heisenberg picture formulation, as
illustrated particularly by the way the unitary operators occur in 
\eqref{jhatdef}, and by the arbitrariness of the surfaces $\Sigma_i^0$,
which may even lie in the future of (some of) the flashes.

\subsection{The Flashes up to Given Surfaces}

It is a corollary of \eqref{ndistr} that the probability for
obtaining, up to spacelike surfaces $\Sigma_i$, a particular sequence
$x_i^1, \ldots, x_i^{n_i}$ of flashes for every type $i$, is
\begin{equation}\label{distruptoSigma}
\begin{split}
  \prob & \Bigl( Q_i \cap \past(\Sigma_i) \in \{x_i^0\}\times d^4x_i^1
  \times \ldots \times d^4x_i^{n_i} \:\: \forall i \Bigr) \: = \\ =&
  \Bigl\langle \Psi^0 \Big| E^{(\vec{n}+1)} \Bigl( \prod_i d^4x_i^1 \times
  \cdots \times d^4x_i^{n_i} \times \fut(\Sigma_i) \Bigr)
  \Big| \Psi^0 \Bigr\rangle
\end{split}
\end{equation}
with $\vec{n}+1 := (n_1+1, \ldots, n_N+1)$.

\subsection{Flash Rate in the Temporal Picture}

It is often desirable to select a time coordinate---a slicing
parametrized by $t$ into spacelike surfaces $\Sigma_i(t)$---in each
$\st_i$ and to employ a picture in which everything evolves as a
function of the time coordinate. In this picture, the interesting
quantity is the probability of having a flash of type\footnote{Unlike as
usual, the capitalization is not meant here to indicate that $I$ is a
random variable, but merely to distinguish that \emph{particular}
number from the other $i$'s.}  $I\in\{1, \ldots,N\}$ between $t$ and
$t+dt$, conditional on the flashes up to time $t$.  For this we obtain
the following formula, in which we assume that the given flashes are
timelike separated, $x_i^k \in \fut(x_i^{k-1})$ for all $i$ and $k
\leq n_i$, and that $d^3y$ is a volume element in $\Sigma_I(t)$. We
denote by $dt \times d^3y$ the 4-volume element between $\Sigma_I(t)$
and $\Sigma_I(t+dt)$ swept out by the normals on $\Sigma_I(t)$ over
$d^3y$; the volume of this element is $c(y) \, dt \, d^3y$ with $c(y)
\, dt = \tdist(\Sigma_I(t+dt),y)$.
\begin{equation}\label{collapserate}
\begin{split}
  \prob & \Bigl( Q_I\cap \fut(\Sigma_I(t)) \cap \past(\Sigma_I(t+dt))
  =\{Y\}, Y\in dt \times d^3y \Big|\\& Q_i \cap \past(\Sigma_i(t)) =
  \{x_i^0, \ldots, x_i^{n_i}\} \; \forall i \Bigr) \: = \\ =& \: c(y)
  \, dt \, d^3y \; \1_{\fut(x_I^{n_I})}(y) \; \frac{1}{c\tau} \exp
  \Bigl( -\frac{1}{c\tau} \tdist(y,x_I^{n_I}) \Bigr) \:\times \\
  &\times \: \frac{ \Bigl\langle \Psi_t \Big|
  E_{1,x_1^{n_1}}^{(1)} \Bigl( \fut(\Sigma_1(t))
  \Bigr) \otimes \cdots \otimes \hat\jmath_I(y)^2 \otimes \cdots \otimes
  E_{N,x_N^{n_N}}^{(1)} \Bigl( \fut(\Sigma_N(t))
  \Bigr) \Big| \Psi_t \Bigr\rangle }{ \Bigl\langle \Psi_t \Big|
  \bigotimes_i E_{i,x_i^{n_i}}^{(1)} \Bigl( \fut(\Sigma_i(t)) \Bigr) \Big|
  \Psi_t \Bigr\rangle }
\end{split}
\end{equation}
where
\begin{equation}\label{Psitdef}
  \Psi_t = \gamma_t \; \Bigl( \bigotimes_i \hat\jmath_i^{n_i} \cdots
  \hat\jmath_i^1 \Bigr) \Psi^0
\end{equation}
with (arbitrary\footnote{That is, although we will consider the usual
normalization $\langle \Psi_t | \Psi_t \rangle =1$,
\eqref{collapserate} holds for any choice of $\gamma_t$, for instance
$\gamma_t=1$. Another useful possibility is to choose $\gamma_t$ such
that the denominator in the last line of \eqref{collapserate} becomes
1.})  normalization factor $\gamma_t$, and $\hat\jmath_I(y)$ (appearing in the $I$-th factor of the tensor product) is the
collapse operator corresponding to a flash at $y$,
\begin{equation}
  \hat\jmath_I(y) =  \hat
  U_{\sph(y, x_I^{n_I})}^{\Sigma_I^0} \: \hat \jmath _{\sph(y,
  x_I^{n_I})} (y) \: \hat U^{\sph(y, x_I^{n_I})}_{\Sigma_I^0} \,.
\end{equation}

\section{The Temporal Picture}\label{sec:temporal}

We now describe the model in the temporal picture, based on a slicing
of space-time.  This is basically a discussion of
\eqref{collapserate}, the flash rate formula of the temporal
picture.

With every time $t$ we associate a wavefunction 
\begin{equation}
  \Psi_t = \gamma_t \; \Bigl( \bigotimes_i \hat{U}^{\Sigma_i(t)}_{\Sigma_i^0}
  \, \hat\jmath_i^{n_i} \cdots
  \hat\jmath_i^1 \Bigr) \Psi^0\,,
\end{equation}
the same as in \eqref{Psitdef} except that we now shift from the
Heisenberg to the Schr\"odinger picture and use the unitary Dirac
evolution to define $\Psi_t$ on $\Sigma_1(t) \times \cdots \times
\Sigma_N(t)$. $\Psi_t$ evolves unitarily apart from collapses.   We remark that
$\Psi_t$ in fact depends only on the surface at time $t$: it
depends on the slicing of space-time only through this surface, in the
sense that different slicings that happen to coincide at $t$ have the
same $\Psi_t$. As is clear from \eqref{Psitdef}, $\Psi_t$ is
determined by the initial wavefunction $\Psi^0$ and the flashes up to
time $t$.

Whenever a flash occurs, the wavefunction $\Psi_t$ gets collapsed by
applying to it the suitable collapse operator $\hat\jmath_i^{n_i+1}$
(and then renormalizing).  Recall that also in the GRW model, the
wavefunction gets collapsed by applying the appropriate collapse
operator.  However, while in the GRW model the collapse operator is
always a multiplication operator in the position representation, this
is not necessarily the case in our model; instead,
$\hat\jmath_i^{n_i+1}$ arises from a multiplication operator on the
suitable hyperboloid, $\sph (X_i^{n_i+1}, X_i^{n_i})$, by using the
Dirac propagators to get to $L^2 (\sph (X_i^{n_i+1}, X_i^{n_i}))$ and
back; in other words, $\hat\jmath_i^{n_i+1}$ is a multiplication
operator evolved, as in the Heisenberg picture, to another surface.

To compute the flash rate, it is unnecessary to know the entire
history of collapses. Instead, it suffices to know the present
wavefunction $\Psi_t$ and the last collapses of all types,
$X_i^{n_i}$.  In other words, the evolution of the $(N+1)$-tuple
$(\Psi_t, X_1^{n_1}, \ldots, X_N^{n_N})$ is a Markov process. Although
this is true as well in the GRW model, there, by way of contrast,
already $\Psi_t$ itself follows a Markov process, and knowledge about
the space-time locations of the previous flashes does not add any
information, beyond what is encoded in $\Psi_t$, about the
distribution of the future flashes.

Another difference between the flash rate formula of our model
\eqref{collapserate} and that of the GRW model
\eqref{grwcollapserate2} is the appearance of a denominator and an
further factors, $E_i^{(1)} \bigl( \fut(\Sigma_i(t)) \bigr)$, in
the numerator; they are needed to correct the rates for conditioning
on the knowledge that no further collapses have occured until time
$t$.

How does the wavefunction transform under a change of slicing of
space-time, such as given by a Lorentz boost in Minkowski space-time? In two
ways. First, some flashes may lie in the future of the new surface
$\Sigma_i'(t)$ but in the past of the old surface $\Sigma_i(t)$ and vice versa;
consequently, the corresponding wavefunctions differ by application of
the collapse operators (respectively their inverses, and
renormalization) belonging to these flashes. Second, on top of that
the wavefunctions differ by the unitary Dirac propagator from one
surface to the other.

\section{The Low Velocity Regime}\label{sec:low}

In this section we consider Minkowski space-time and assume the
existence of a Lorentz frame in which all velocities, at which wave
packets move or spread, are small compared to the speed of light. We
point out that in this regime, and in this frame, our model approaches
the GRW model.

The basic observation is that for large $t$, such as $t$ of the order
of magnitude of $\tau \approx 10^{15}\, \text{sec}$, the hyperboloid
$\sph_{ct}$ becomes rather flat and is well approximated by the plane
$x^0=ct$ tangent to the hyperboloid, at least in the neighborhood of
their common point $(ct,0,0,0)$ with radius $R$ such that $R \ll ct$,
or $R \ll 10^8 \: \text{light-years}$.  As a consequence, to multiply
by a Gaussian with width $a \ll R$ centered somewhere inside this
neighborhood means approximately the same on the hyperboloid and on
the plane. And since, by the assumption on velocities, the wavefunction
will lie almost completely within this neighborhood for $t$ of the
order of magnitude of $\tau$, the action of the collapse operator
agrees approximately with that of the GRW model. Similarly, the factor
$\1_{\fut(X)}$ is 1 on the support of $\Psi_t$.

Replacing our collapse operators with those of GRW, we obtain
$E_i^{(1)} \bigl( \fut(\Sigma_i(t)) \bigr) \approx e^{-(t-T_i)/\tau}
\: \hat{1}$, so that, since $c(y) = c$ for this slicing,
\eqref{collapserate} in fact reduces to \eqref{grwcollapserate2}.

There is a subtlety in that if $N$ is very large, flashes do sometimes
occur for which the timelike distance $c\Delta T$ from the previous
flash of the same type is not of the order of magnitude of $c\tau$,
but much smaller.  (However, $c\Delta T$ is still much larger than $a$
except in very rare cases that occur only once in $c\tau/a \approx 
10^{30}$ collapses.) Fortunately, relevant wave functions obey a bound
on their spread: it is at most $a$ initially (thanks to the previous collapse) and grows much slower than at rate $c$. Thus, the spread is still much smaller
than $c \Delta T$, so that the collapse operators of the two
models do not differ appreciably on these wavefunctions.

\section{Predictions}\label{sec:predictions}

What predictions does the model entail? To what extent are they in
agreement with quantum mechanics? In what way do they
deviate from the predictions of the nonrelativistic GRW model? 

To begin with, a difficulty with obtaining any predictions at all from
the model is that it does not involve any interaction, and thus does not
support the formation of macroscopic bodies such as 
observers or apparatuses.  However,
we can say what the predictions will be like once interaction is
included in whatever way, be it by particle creation and annihilation
or by a modification of the unitary propagators: we can consider a
wavefunction as would arise from interaction.

As a corollary of the previous section, for any experiment in which no
parts need move at relativistic speeds our model approximates the GRW
model and is thus in agreement with quantum mechanics to the same
extent as the GRW model; this includes all presently doable
experiments; for a discussion of future experiments that may
distinguish spontaneous collapse theories from quantum mechanics, see
Section~V of \cite{overview}.  The model is in particular in agreement
with the result of EPR--Bell experiments such as Aspect's
\cite{aspect} and thus violates Bell's locality inequality
\cite{Bellbook}.

A general pattern of behavior of the model follows from another trait
it has in common with the GRW model: that disentangled subsystems are
governed by the same laws as the whole, and follow an independent
collapse process. Therefore, a small system of $N_1 < 10^5$ particles
will not collapse for the next thousand years, provided it stays
disentangled.  A macroscopic body, however, cannot support
superpositions over distances much wider than $a \approx
10^{-7}\, \text{m}$ for longer than a split second.

The most obvious deviation of our model from the GRW model is that a
system moving at a speed close to $c$ will have a reduced rate of
spontaneous collapse, reduced by just the factor that one would expect
from a naive application of time dilation.  This will be hard to see
in experiment, of course, given that as yet we cannot see any
spontaneous collapses in experiment.

Extensions of the model to quantum field theory might deviate from the nonrelativistic GRW model in a prediction of spontaneous pair creation at the flashes, as
discussed in Section~\ref{sec:positron}.

Superluminal communication on the basis of entangled sets of particles
is impossible in our model.  To see this, suppose that Alice and Bob
are widely separated and share a system of entangled particles; the
marginal distribution of the $Q_i$ for the $i$'s of all particles
located on Alice's side is, as follows from \eqref{ndistr} and
\eqref{Eidef}, independent of the fields applied to the $x_i$ for the
$i$'s of the particles on Bob's side, such as the metric and
electromagnetic vector potential of $\st_i$.

\section{Perspective}\label{sec:conclusions}

Our model seems to be the first model in the literature that achieves
all of the following: 
\begin{itemize}
\item[(i)] it describes a possible (many-particle)
world in which outcomes of experiments have (to a sufficient degree of
accuracy for all cases presently testable) the probabilities
prescribed by quantum theory, 
\item[(ii)] it describes objective events in space-time (the flashes), in contrast
to theories merely associating a wavefunction with every spacelike surface,
\item[(iii)] it is fully compatible with
relativity in that it does not rely on a preferred slicing of
space-time, and 
\item[(iv)] it works in the continuum, in contrast to theories assuming 
a discrete space-time. 
\end{itemize}

We give a brief overview of the literature concerning relativistic
models explaining the probabilities prescribed by quantum theory. Such
models come in two varieties, either as a variant of Bohmian mechanics
or as a spontaneous collapse theory.

Among the variants of Bohmian mechanics, Bohm~\cite{Bohm53} gives a
Lorentz-invariant equation of motion for a single Dirac particle; Bohm
and Hiley~\cite{BH} give a many-particle version based on a preferred
Lorentz frame; D\"urr et al.~\cite{HBD} generalize to an arbitrary
spacelike preferred slicing of space-time, possibly determined by a
covariant law involving the wavefunction; Samols~\cite{samols1,samols2} 
gives a Bohm-type model
on a discrete space-time using a preferred slicing; Goldstein and
Tumulka~\cite{arrows} give a nonlocal many-particle version without
preferred slicing, which however fails to yield any probabilities.
Both Berndl et al.~\cite{berndl} and Dewdney and
Horton~\cite{synchro1} suggest, instead of a preferred slicing, a
preferred joint parametrization (or synchronization) of the world
lines, and thus obtain a nonlocal Bohm-type dynamics; however, this
does not really conform any better with the spirit of relativity than
a preferred slicing; in addition, the models fail to yield any
probabilities.  

All spontaneous collapse models deviate slightly from the quantum
prescriptions in their probabilities. An overview of spontaneous
collapse models is given by Bassi and Ghirardi in \cite{overview}.
Dove and Squires~\cite{dovethesis,dovepaper} have made steps towards a
relativistic model based on discrete flashes.\footnote{Their paper
\cite{dovepaper} (reprinted, with minor extensions, as Chapter~7 of
\cite{dovethesis}) is irritating in that the authors claim to provide
a Lorentz-invariant collapse model but do not keep their promise. They
define an evolution of the wavefunction \emph{given} the flashes,
according to which the wavefunction collapses on the future light cone
of each flash and evolves unitarily in between. However, they do not
specify a probability law for the flashes.}  Dowker and
coworkers~\cite{dowker1,dowker2} give a collapse model on a discrete
space-time that does not need a preferred slicing; it is not known how
this model could be adapted to a continuum.  All other efforts towards
a relativistic collapse model are based on the approach of continuous
spontaneous localization (CSL), corresponding to diffusion processes
in Hilbert space; some references describing research in this
direction are \cite{pearle90, 67, diosi90, pearle99, NR}.

A somewhat surprising feature of the present situation is that we seem
to arrive at the following alternative: Bohmian mechanics shows that
one can explain quantum mechanics, exactly and completely, if one is
willing to pay the price of using a preferred slicing of space-time; our model
suggests that one should be able to avoid a preferred slicing if one is
willing to pay the price of a certain deviation from quantum mechanics.

\bigskip

\noindent \textit{Note added.} Several articles discussing the relativistic model
of this paper have been written since its first preprint version 
became available in 2004 \cite{Mau05,Tum06,AGTZ06}.

\bigskip

\noindent \textit{Acknowledgments.} I wish to thank Fay Dowker, Detlef
D\"urr, Shelly Goldstein, Philip Pearle, and Nino Zangh\`\i\ for their
critical comments on a draft.  I have also profited a lot from
discussions with each of them and, in addition, with GianCarlo
Ghirardi and Peter Pickl. This work was supported in part by INFN.

\end{document}